\documentclass[sigconf]{acmart}

\AtBeginDocument{%
  }

\copyrightyear{2025}
\acmYear{2025}
\setcopyright{cc}
\setcctype[4.0]{by}
\acmConference[IUI '25]{30th International Conference on Intelligent User Interfaces}{March 24--27, 2025}{Cagliari, Italy}
\acmBooktitle{30th International Conference on Intelligent User Interfaces (IUI '25), March 24--27, 2025, Cagliari, Italy}\acmDOI{10.1145/3708359.3712144}
\acmISBN{979-8-4007-1306-4/25/03}

\usepackage[english]{babel}

\usepackage{enumitem}
\usepackage{listings}
\usepackage{multirow}

\newcommand{\rr}[1]{\textcolor{black}{{#1}}}

\begin{document}

\title{\sysname{}: Aggregating How-To Videos for Task-Oriented Learning}


\author{Saelyne Yang}
\email{saelyne@kaist.ac.kr}
\authornote{This work was done while the author was an intern at Adobe Research.}
\affiliation{%
  \institution{School of Computing, KAIST}
  \city{Daejeon}
  \country{Republic of Korea}
}

\author{Anh Truong}
\email{truong@adobe.com}
\affiliation{%
  \institution{Adobe Research, USA}
  \city{New York, NY}
  \country{USA}
}

\author{Juho Kim}
\email{juhokim@kaist.ac.kr}
\affiliation{%
  \institution{School of Computing, KAIST}
  \city{Daejeon}
  \country{Republic of Korea}
}

\author{Dingzeyu Li}
\email{dinli@adobe.com}
\affiliation{%
  \institution{Adobe Research, USA}
  \city{Seattle, WA}
  \country{USA}
}

\renewcommand{\shortauthors}{Saelyne Yang, Anh Truong, Juho Kim, Dingzeyu Li}
\newcommand{\sysname}{VideoMix}

\begin{abstract}
Tutorial videos are a valuable resource for people looking to learn new tasks. People often learn these skills by viewing multiple tutorial videos to get an overall understanding of a task by looking at different approaches to achieve the task. However, navigating through multiple videos can be time-consuming and mentally demanding as these videos are scattered and not easy to skim.
We propose VideoMix, a system that helps users gain a holistic understanding of a how-to task by aggregating information from multiple videos on the task. Insights from our formative study (N=12) reveal that learners value understanding potential outcomes, required materials, alternative methods, and important details shared by different videos. Powered by a Vision-Language Model pipeline, VideoMix extracts and organizes this information, presenting concise textual summaries alongside relevant video clips, enabling users to quickly digest and navigate the content.
A comparative user study (N=12) demonstrated that VideoMix enabled participants to gain a more comprehensive understanding of tasks with greater efficiency than a baseline video interface, where videos are viewed independently. Our findings highlight the potential of a task-oriented, multi-video approach where videos are organized around a shared goal, offering an enhanced alternative to conventional video-based learning.

\end{abstract}




\begin{teaserfigure}
  \includegraphics[width=\textwidth]{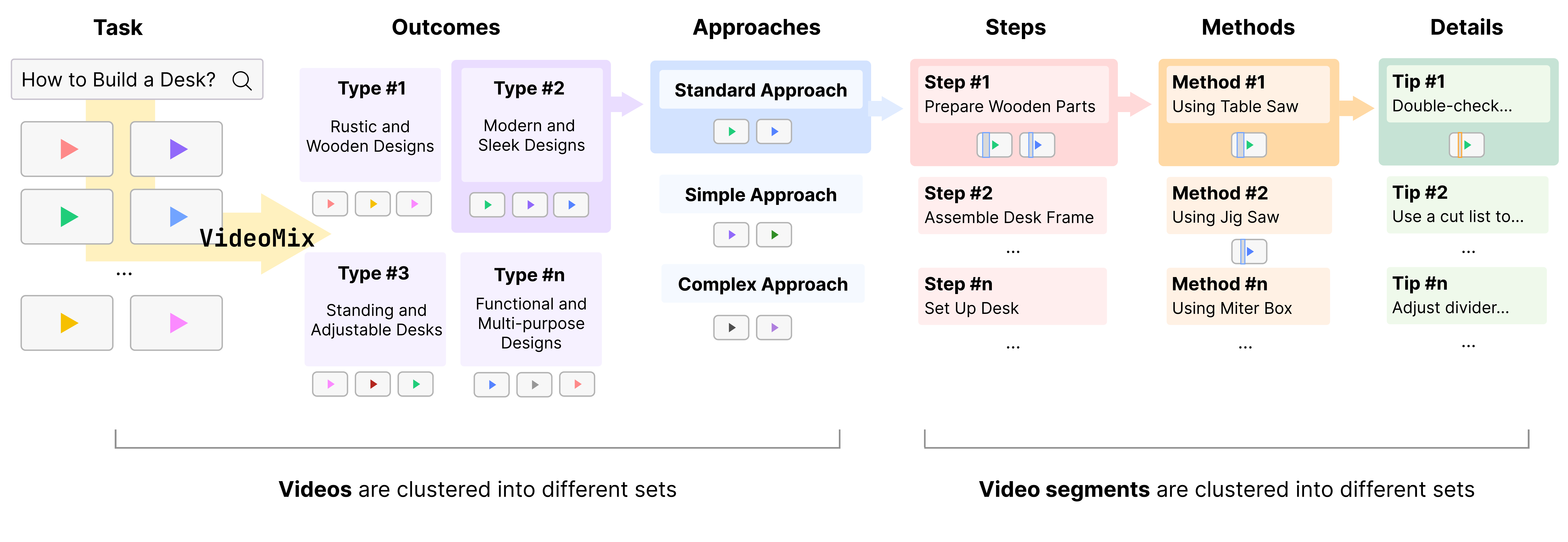}
  \caption{\sysname{}'s process for organizing information from multiple how-to videos. Given a set of videos on the same task, it first identifies the different types of outcomes. For each outcome type, \sysname{} determines standard, simple, and complex approaches to achieve the desired result, from the videos that correspond to that outcome type. 
  Once a video set is determined that follows each approach, \sysname{} identifies parts of videos that correspond to each step in the approach. For each step, it identifies different methods employed in the videos, along with additional details such as tips and notes for each method.}
  \label{fig:teaser}
  \Description{A diagram that describes the process of organizing information from multiple videos; From task to outcomes, approaches, steps, methods, and details.}
\end{teaserfigure}


\maketitle

\section{Introduction}

How-to videos are a popular resource for people looking to learn new tasks (e.g., cooking a pasta dish or knitting a mitten) due to their abundance and the detailed, step-by-step instructions~\cite{yang23howtotaxonomy, chi2012MixT}.
\rr{When learning a task, people typically start with understanding the procedure and then applying it in their specific context~\cite{choi23procedural, anderson2001taxonomy, yang24skills}. This process involves gathering and processing information to construct an understanding of the task, followed by active engagement through execution and iterative learning via trial and error.}

\rr{In the initial phase of learning,} people often develop their understanding by watching or skimming through \textit{multiple} videos.
Watching multiple videos on the same topic can significantly enhance understanding of the task, by offering diverse perspectives and insights~\cite{community-enhanced-tutorial, kim23surch}.
This exposure allows users to learn about different methods, tips, or prerequisites, and select the approach that best fits their context. Additionally, learners can reference different videos to clarify any unclear points or to confirm the reliability of a specific method.

\rr{While this diversity provides such benefits, making sense of the loads of information in multiple videos is challenging.} These videos are not curated, leaving the job of organizing and tailoring the information for the personal needs on the user. 
Navigating through numerous videos can be time-consuming, as most platforms are designed for viewing one video at a time, making related content fragmented and scattered. 
Moreover, since videos are not easy to skim, users must watch them sequentially, which can be inefficient. As a result, learners may end up watching only a few, potentially missing out on valuable information and knowledge. 
While systems like Surch~\cite{kim23surch} and RecipeScape~\cite{chang2018recipescape} aggregate multiple procedures for a common task, they are specialized for specific domains (e.g., surgery) or primarily designed for analytical purposes, which often require domain expertise. Further exploration is needed to support learners in building a well-rounded understanding of tasks across a variety of domains.

To better understand why users watch multiple videos and what specific information they seek to gain from this process, we conducted a formative study in which we asked twelve participants to learn a task of their choice using how-to videos. We found that learners primarily look for four key aspects in the videos: \textbf{1) Outcomes}, to understand the possible results of the task and decide which outcome they prefer; \textbf{2) Requirements}, to identify the necessary tools or materials, and check whether certain tools are commonly used across videos; \textbf{3) Approaches and Methods}, to explore alternative approaches presented by various instructors and find the method that best suits their needs; \textbf{4) Details}, to gather additional insights, such as tips or know-how shared by different instructors. While participants recognized the value of watching multiple videos to gather this information, they noted the difficulty of tracking and organizing the information and the inefficiency of navigating between multiple videos.

Based on these findings, we developed \sysname{}, a system that aggregates and organizes information from multiple how-to videos on a single task, helping users gain a holistic understanding of the task. 
\sysname{} focuses on physical tasks with tangible outcomes, \rr{organizing videos into meaningful axes; outcomes, approaches, steps, methods, and details (Figure~\ref{fig:teaser}).}
Once the user inputs a task they want to learn, 
\sysname{} identifies different outcome types (Figure~\ref{fig:system-overview}B), and for each outcome type, \sysname{} provides three different approaches to achieve the outcome: the standard (most commonly followed), the simplest (with the fewest steps), and the most complex (with additional steps) approaches (Figure~\ref{fig:system-overview}C). Each approach is presented with the specific steps that make up the process, accompanied by a list of materials and tools used across the videos (Figure~\ref{fig:system-overview}D, E).
Once the user selects an approach they are interested in, they can explore different methods to achieve each step (Figure~\ref{fig:system-detail}B). \sysname{} provides video snippets demonstrating each method, along with useful tips or important details drawn from the videos (Figure~\ref{fig:system-overview}C, E). To present potentially heterogeneous information from multiple videos in a coherent and digestable way, we integrate concise textual summaries with relevant video clips, enabling users to quickly digest and navigate the content.

To extract and generate this information, we designed a technical pipeline powered by a Vision-Language Model (VLM). Our pipeline processes a collection of videos to automatically extract key information such as outcome types, requirements, and step information along with relevant details from both the visual and verbal content of videos. A key component of our pipeline is the Dynamic Approach Identification (DAI) module, which captures different possible sequences of steps to achieve an intended outcome from a set of videos.

To evaluate \sysname{}, we conducted a within-subjects study (N=12), where participants were asked to learn tasks that they had not done before, with our system and a conventional YouTube-like system. The results revealed that \sysname{} helped participants gain an overall understanding of the task more efficiently, allowing them to tailor their learning experience by exploring approaches that matched their interests and suited their needs. Overall, \sysname{} demonstrates the potential of task-based learning for videos, where videos are organized around a common task or goal, offering a concise yet comprehensive resource.

This paper presents the following contributions:
\begin{itemize}
    \item{A formative study that uncovers how users learn from multiple videos.}
    \item{\sysname{}, a system that aggregates and presents information from multiple how-to videos on a task.}
    \item{An evaluation study that demonstrates the effectiveness of our system in task learning.}
\end{itemize}


\section{Related Work}
Our work presents a system that aggregates information from multiple how-to videos to assist users in task learning. We review related work on how-to video learning, improving video skimming and browsing, and handling multiple instructions and workflows.

\subsection{How-to Video Learning}
Enhancing learning from how-to videos has been extensively explored in HCI research. A common approach is to segment videos into meaningful units, making it easier for learners to process information~\cite{chi2012MixT, truong2021makeup}. These segments can be tutorial steps~\cite{fraser2020temporal, chi2012MixT, truong2021makeup}, types of information~\cite{yang23howtotaxonomy}, or intermediate outcomes~\cite{nawhal2019videowhiz, kim2014toolscape}. Such segmentation helps users with video navigation and comprehension, facilitating their learning experience. 

Other research has focused on supporting learners in following along with tutorial videos. In digital how-to tasks, learners often struggle to apply what they see in software tutorials due to context mismatches, such as interface differences between the video and their application~\cite{hardFollowTutorial, tutorialMismatch}. To address this, systems like ReMap~\cite{fraser20remap} and Replay~\cite{fraser19replay} have been proposed to help users locate specific video segments relevant to their query and software context. SoftVideo~\cite{yang2022softvideo} and Pause-and-Play~\cite{Pongnumkul2011pause} take this further by tracking users’ progress in both the tutorial video and the software application.
For physical how-to tasks, where users' hands are often occupied (e.g., using tools while watching), several studies have explored voice-based video control~\cite{chang2019howto, zhao22jiggling, lin23context}. They identify various challenges in conversational interaction with videos. RubySlippers~\cite{chang2021rubyslippers} addresses some of these issues by enabling keyword-based voice navigation, making it easier for users to control video when they follow the tutorial in their context.

While many of these studies focus on extracting and surfacing key information components such as steps, they often address a single video. Our approach extends this by drawing information from multiple videos. This allows us to support broader learning, where users aim to understand a task holistically rather than learning from just one video. Additionally, unlike previous work that centers on helping users follow along with videos, we focus on learning by watching, where users seek to gain a comprehensive overview of the task.

\subsection{Making Videos Skimmable}
When presenting information from multiple videos, it is important to organize the content in a structured manner to avoid overwhelming users. In our system, multi-video information is provided progressively, allowing users to select their preferred learning paths.  Prior work has explored improving the browsing of multiple video snippets by organizing frames along meaningful dimensions for video editing~\cite{lin24videomap}, or content exploration~\cite{Matejka2014videolens, zhang14videograph}. However, these approaches typically focus on visual frames, sorting them in latent space, or rely on metadata for a specific application.
In instructional how-to videos, however, verbal content also carries critical information~\cite{yang23howtotaxonomy, yang2024ytcommentqa}, adding an additional layer of complexity. These videos often contain a richer depth of knowledge, delivered through both visual and verbal channels. Building on these ideas, we enhance multi-video skimming specifically for how-to videos, where users must process and synthesize more complex, detailed information from multiple sources.

We also present information in mixed-media formats—combining text, images, and video snippets— to ensure the content is easy to browse and skim. Prior work has explored the use of mixed-media formats to enhance video skimming~\cite{pavel2014digests, pavel2015sceneskim}, showing their effectiveness in how-to learning scenarios~\cite{chi2012MixT, truong2021makeup}. Several systems have focused on authoring tutorials by combining appropriate assets like text, images, and videos~\cite{chi2021markdown, recipe2video}, with TutoAI~\cite{chen2024tutoai} proposing a framework for creating mixed-media tutorials. However, these systems typically focus on a single video at a time. We build upon these ideas to support users in skimming and exploring content across multiple videos simultaneously.

\subsection{Multiple Instructions and Workflows}
Learning from multiple resources can foster a deeper understanding of a subject~\cite{learningmultipletext, undergradMultiple}. For example, FollowUs~\cite{community-enhanced-tutorial} demonstrated the effectiveness of offering multiple demonstrations of a tutorial performed by different users, providing various insights and allowing learners to pick up on pieces from different tutorials.  
To facilitate this, researchers have developed systems that enable the comparison of hundreds of cooking recipes~\cite{chang2018recipescape} or software workflows~\cite{kong12delta}, as well as computational pipelines that capture the diversity of these demonstrations~\cite{chang2020workflow, wang2018command}. 
A similar approach has been explored in the context of multi-document analysis, where systems were proposed to effectively collect and organize information from multiple relevant documents~\cite{fok24marco, han22passages}. 
In video-specific research, several systems have been proposed to facilitate multi-video analysis. For example, Surch~\cite{kim23surch} enables structured search and comparison of surgical videos, while Video Lens~\cite{Matejka2014videolens} offers interactive search and exploration of baseball videos.
Recently, the Computer Vision community has developed methods to detect differences in the same instructional step between two videos~\cite{stepdiff} or to navigate to a video that demonstrates a different approach for the same step~\cite{ashutosh2024detours}. Our research builds on this prior work by aggregating and organizing hundreds of videos across a wide range of instructional domains, providing a comprehensive, bird's-eye view of a task.
By doing so, we aim to achieve the benefits of learning from multiple resources in video-based learning.

\section{Formative Study}

We conducted a formative study to gain insights into how users learn new tasks through multiple how-to videos and to understand the specific information they seek across these videos. In this section, we describe the methodology used and key findings identified from the study.

\subsection{Method}
We recruited 12 participants (6 male, 6 female, \rr{mean age=27.7, median=27}) through online
communities of academic institutions, who regularly watch how-to videos and often watch multiple videos to gain a comprehensive understanding of a task. All participants reported that they watch how-to videos of various domains such as cooking, painting, gardening, and assembly, at least 1-2 times per month.

To begin, we asked participants a few questions about their current practices on learning from how-to videos. We asked about the types of how-to videos they usually watch and asked them to describe their typical workflow, from watching the videos to following through with the task.

Next, participants were asked to select a topic or task they wanted to learn, ensuring it was a subject they had not previously learned or explored. Once the task was chosen, we conducted a think-aloud observation study. Participants were instructed to open YouTube, share their screen, and learn about the selected task as they would normally do. 
To simulate a realistic learning scenario, we asked them to imagine a setting where they had to learn about the task so that they could execute the task later. During the session, we observed how participants searched for videos, the specific information they sought, when and why they chose to look for another video and switch between them, and what information they gathered from each video. Participants were encouraged to think aloud about their thought process throughout the learning phase. We repeated the observation study with at least two tasks of the participant's choice, within a 45-minute timeframe.

Following the observation study, we conducted a semi-structured interview. We asked participants to describe the overall approach they used to learn the task, the types of information they found useful from different videos, the challenges they encountered, and the kind of support they would find helpful when navigating through multiple videos. The study was conducted online, and participants were compensated with a \$30 USD Amazon gift card for the 1-hour session.

\subsection{Findings}


\subsubsection{Current Workflows}
\rr{All participants mentioned that when learning a task, they typically start by watching videos to understand the materials, processes, and techniques involved, forming a mental map before following the task.
To watch videos, all participants began their video search with broad, general queries (e.g., \textit{`how to make gnocchi'}), believing} that these general queries would provide a better overview of the task and increase the chances of finding higher-quality videos, as a larger video pool is more likely to contain quality content.
In contrast, they believed that more specific queries with personal contexts or constraints (e.g., \textit{`how to make gnocchi without potato'}) might limit the search results. Additionally, since participants did not yet have an understanding of the task, they were often unsure about what specific details would be relevant to include in the search.

These broad queries yielded a large number of videos. All participants watched multiple videos when learning the task, and demonstrated two common behaviors for navigating through them. In the first behavior, \rr{demonstrated by five participants,} they quickly scanned a list of videos and opened several videos in separate tabs, selecting those that aligned with their interests based on factors such as an appealing outcome, a title that matched their expectations (e.g., `simple recipe'), or visual cues suggesting the video was of high quality.
In the second behavior\rr{, observed in seven participants,} they selected one video to watch at a time. Through watching that video, participants developed a better understanding of which personal constraints were relevant (e.g., not having a tool they needed), what specific outcome they wanted, or any knowledge gaps they needed to be clarified. They then accordingly refined their search queries for subsequent videos to become more specific and tailored to those needs.

\subsubsection{Information Users Expect to See from Multiple Videos}
\label{section:information-types}

Watching multiple videos allowed participants to get a broader understanding of the task and see various approaches and details that might not be covered in a single video. Below are the key pieces of information participants sought from multiple sources:

\textbf{Outcomes}: Participants quickly scanned video thumbnails and titles to grasp the specific outcomes of the task. For example, in learning how to "make gnocchi," they encountered variations like "cream gnocchi," "basil gnocchi," or "gnocchi soups." This allowed them to compare different end results and decide which version they wanted to pursue. Understanding these possible outcomes helped participants shape their goals and choose the appropriate approach.

\textbf{Requirements}: Participants also looked for the tools, materials, or ingredients used in the videos. By observing the requirements across multiple videos, they could identify commonly used items and ensure they had everything necessary to complete the task. This also allowed them to compare any unique items suggested by different instructors, helping them decide which tools or materials were essential.

\textbf{Approaches and Methods}:  Participants explored various workflows presented in the videos, helping them identify both standard and alternative approaches. This comparison allowed them to understand the complexity of different methods and select one that best matched their skill level or specific context. Additionally, learning about different alternative methods provided flexibility and adaptability in their learning process.

\textbf{Details}: Participants appreciated the additional details that different videos provided, such as tips, tricks, or know-how. These insights added value to the learning experience, giving them more in-depth or practical knowledge that could enhance their understanding of the task.

\subsubsection{Challenges}

While participants found that watching multiple videos to be very beneficial to their learning, they also noted that the current process for using multiple videos is time-consuming and mentally demanding. 
They encountered the following challenges while trying to select, watch and organize information from multiple videos:

\textbf{Search Results Lack Organization:}
The search queries always returned a large number of videos that weren't organized in a way participants could understand. As a result, participants found it difficult to select which video or videos to watch from the large set. For example, all participants primarily selected videos based on the outcome, which they determined from the search result titles and thumbnails. However, the search results were not organized by outcome; videos sharing a common outcome were scattered throughout the result list and participants had to exhaustively examine the list to comprehend all the possible outcomes for the task. Moreover, it was difficult for participants to gauge how videos sharing a common outcome differed. Better organization of the task videos based on the expected information types (Section \ref{section:information-types}) could help to reduce users' mental load.

\textbf{Information Extraction Requires Watching Videos:}
Participants found it difficult to skim videos and spent a significant amount of time watching each video end-to-end in order to extract the information they wanted. For example, unless the original creators manually annotated the video or specified in the description box, participants often had no quick way to determine all the steps or ingredients used without watching the video through and risked missing important information while skimming. In contrast to video-only interfaces, past research has shown that mixed-media tutorials, which incorporate text, images and video together, are easier to skim and more effective at giving users a high-level overview of the task \cite{chi2012MixT, truong2021makeup}.

\textbf{No Easy Way to Compare and Consolidate Information Across Videos:}
As participants watch multiples video, they don't just want to gather information about each video independently. Instead, they were trying to form broader task insights which span multiple videos such as what the common approach is, which steps are not strictly necessary, or different methods to execute a single step. However, current video interfaces only support single video contexts; in order to watch multiple videos, participants had to open each video in a new tab and the videos were not aligned to each other in any way. This interface design made it difficult for participants to compare multiple videos and, as a result, participants spent considerable mental effort synthesizing and tracking these task insights. Additionally, participants also wanted to aggregate information across videos (e.g., all the tips and details from different instructors about a single step), but had no way of doing so in the current video browsing interface. Multiple participants expressed a desire for a system that could help them connect and consolidate the information from multiple videos more effectively.


\subsection{Design Goals}
From the formative study, we observed that watching multiple videos offered participants a more comprehensive understanding of a task, enriched with diverse instructions and insights. However, there was a need for a more efficient way to access and organize this information.
Based on the study insights, we derive the following design goals for a multi-video system that is designed around a common task goal:
\begin{itemize}
    \item DG1: Enable users to gain a comprehensive overview of possible outcomes and requirements for the task.
    \item DG2: Help users compare and navigate different approaches and methods to achieve the task.
    \item DG3: Provide easy access to detailed information, including relevant video snippets and key details shared across multiple videos.
\end{itemize}
\section{VideoMix}

Based on our design goals, we present \sysname{}, a system that helps users gain a holistic understanding of a how-to task, by aggregating and organizing information extracted from multiple videos on the task.

\subsection{System Interface}
The system consists of an (1) Overview page (Figure~\ref{fig:system-overview}) and (2) Details page (Figure~\ref{fig:system-detail}). The overview page gives an overview of the task by organizing possible outcomes of the task, required materials and tools, and several approaches to achieve the task. Once the user selects an approach they are interested in, they see the steps that the approach involves. Once they click on a step, the system takes the user to the Details page, where users can see details for the step including multiple alternative methods and important tips, along with the corresponding video snippets. 

\subsubsection{Overview page}
\begin{figure*}[t]
  \includegraphics[width=0.9\linewidth]{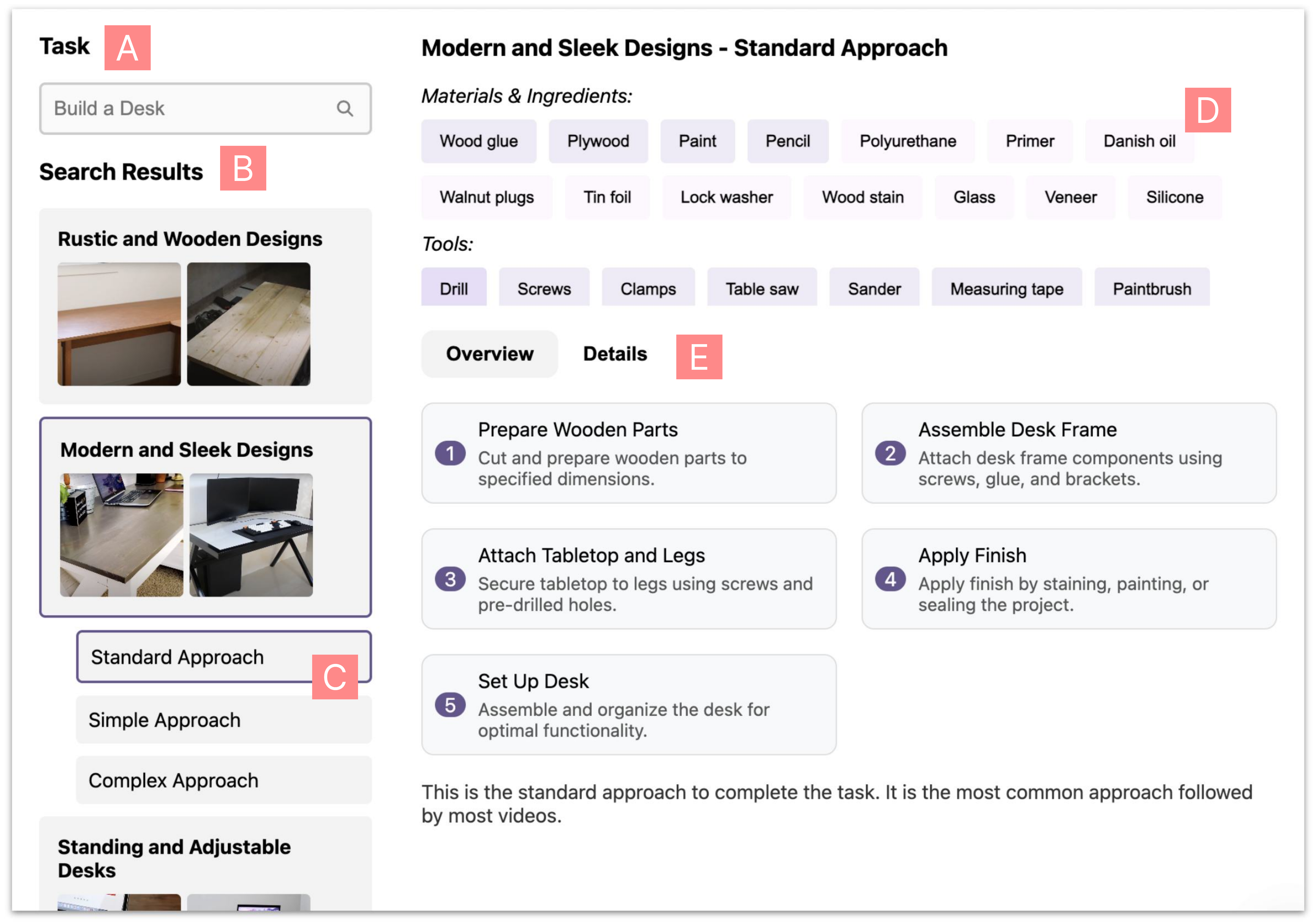}
  \caption{\sysname{} interface on the Overview page for the task ``Build a Desk’’. (A) Users begin by selecting the task they want to learn. (B) \sysname{} then presents video search results categorized by outcome types. (C) For each outcome type, users can choose from standard, simple, or complex approaches. (D) Based on the chosen approach, \sysname{} displays the necessary requirements, such as materials, ingredients, and tools. Finally, (E) users can see a list of steps and a brief description of each step that makes up the chosen approach.\protect\footnotemark}
  \label{fig:system-overview}
  \Description{The Overview page of VideoMix. There is a task name and search results organized by outcome types and approaches on the left, and materials and tools and step information on the right.}
\end{figure*}

Once the user specifies the task they want to learn, \sysname{} presents an overview of that task. First, it offers several possible outcomes (Figure~\ref{fig:system-overview}B) for the task (e.g., for the task "Build a Desk," it shows options like "Rustic Wooden Design," "Modern Sleek Design," "Functional Multi-purpose Desk," or "Standing Adjustable Desk"). 

After the user selects a preferred outcome, \sysname{} provides three different approaches (Figure~\ref{fig:system-overview}C) to achieve it: the standard approach (the most commonly used across videos), the simplest approach (involving the fewest steps), and the most complex approach (involving the most steps).
These approaches inform users of multiple ways to accomplish the task, varying in both commonness and complexity, while also providing flexible options tailored to their experience level and the amount of effort they wish to invest.

Once the user selects an approach, the system provides an overview of information gathered from multiple videos corresponding to that approach. First, a list of materials and tools used in the videos that follow the approach is provided (Figure~\ref{fig:system-overview}D). Since not all items are used in every video, they are sorted by frequency of use—items appearing more often are highlighted with darker colors, making it easy for users to identify the most commonly used ones. Below the item list, the system displays step-by-step information for the approach, with each step labeled and briefly described (Figure~\ref{fig:system-overview}E). 


\subsubsection{Details Page}
Once the user selects a step in an approach, they are presented with more in-depth information on the Details page(Figure~\ref{fig:system-detail}). In this detailed view, \sysname{} displays the step-by-step instructions previously shown in the Overview page, in a vertical format (Figure~\ref{fig:system-detail}A). Here, each step can be expanded to reveal multiple variations or methods for accomplishing that step (Figure~\ref{fig:system-detail}B). For example, for the step "Cook meat and vegetables," the user can choose between methods such as "Using an Instant Pot," "Using a Rice Cooker," or "Using a Cast Iron Pot." 

Once the user selects a method, \sysname{} presents video snippets corresponding to the chosen method (Figure~\ref{fig:system-detail}C). These videos automatically play from the relevant start time and stop at the end of the segment, but users have the option to explore the video further by watching earlier or later parts to understand its context. On the right side of the video player, users can navigate between different video snippets, each accompanied by a brief summary (Figure~\ref{fig:system-detail}D). This allows users to quickly understand the content of each snippet before selecting one to view, helping them explore different videos demonstrating the method. 
Below the video player, \sysname{} provides useful tips and key information extracted from the video snippets to highlight important points or considerations for the selected method (Figure~\ref{fig:system-detail}E).

As such, \sysname{} enables users to gain a comprehensive understanding of the task by presenting information in a structured and hierarchical manner. This approach allows users to progressively learn about the task, revealing details as they delve into each outcome, approach, and step in depth.



\subsection{Technical Pipeline}

To provide the aggregated information from multiple how-to videos, we developed a pipeline that processes and extracts content in videos. Figure~\ref{fig:teaser} illustrates the overall process. It begins by clustering videos into different sets based on their outcome and approach type. Each video set is then analyzed to extract more detailed information, such as steps and methods used.
For the video dataset, we used HowTo100M~\cite{miech19howto100m}, a large-scale collection of narrated how-to videos from YouTube. 
\rr{We downloaded the corresponding YouTube videos using youtube-dl~\cite{youtubedl}, a command-line program for downloading videos from YouTube.
We then obtained the video transcripts open-sourced by Han et al.~\cite{han2022align}, which were generated with sentence-level timestamps using WhisperX~\cite{bain2022whisperx}.}
Each video is labeled with its task name (e.g., `make gnocchi'), along with a broader category it belongs (e.g., `Food and Entertaining').

\footnotetext{\rr{Screenshots of the outcome search results are from: \href{https://youtu.be/CbJtZFXwxKY}{youtu.be/CbJtZFXwxKY}, 
\href{https://youtu.be/Fnl1OwAAvEo}{youtu.be/Fnl1OwAAvEo}, 
\href{https://youtu.be/Z7x_Rvb_yjc}{youtu.be/Z7x\_Rvb\_yjc}, 
\href{https://youtu.be/_v0fXgwcrpY}{youtu.be/\_v0fXgwcrpY} (Creative Commons licensed)
}.}

\subsubsection{Outcomes}
To determine the different outcome types for a task, our pipeline operates in two phases: first, it extracts descriptions of each video’s outcome and then it clusters these outcome descriptions into meaningful categories. In the first phase, we utilize both the visual content and transcripts. While transcripts provide verbal descriptions of the outcome~\cite{yang23howtotaxonomy}, visuals can offer additional descriptive information that may not be explicitly mentioned. To estimate which video frames show the outcome, we provide GPT-4o with the full transcript and prompt it to extract only the segments that describe the outcome (see Appendix~\ref{sec:prompts_outcome_transcript}). We pick the video frames that correspond to these transcript segments as outcome frames, \rr{selecting one frame per second}. We then input these outcome frames and the entire transcript into GPT-4o and prompt it to generate an outcome description (Appendix~\ref{sec:prompts_outcome_description}). This phase yields an outcome description for each video in the task set.

\begin{figure*}[t]
  \includegraphics[width=0.9\linewidth]{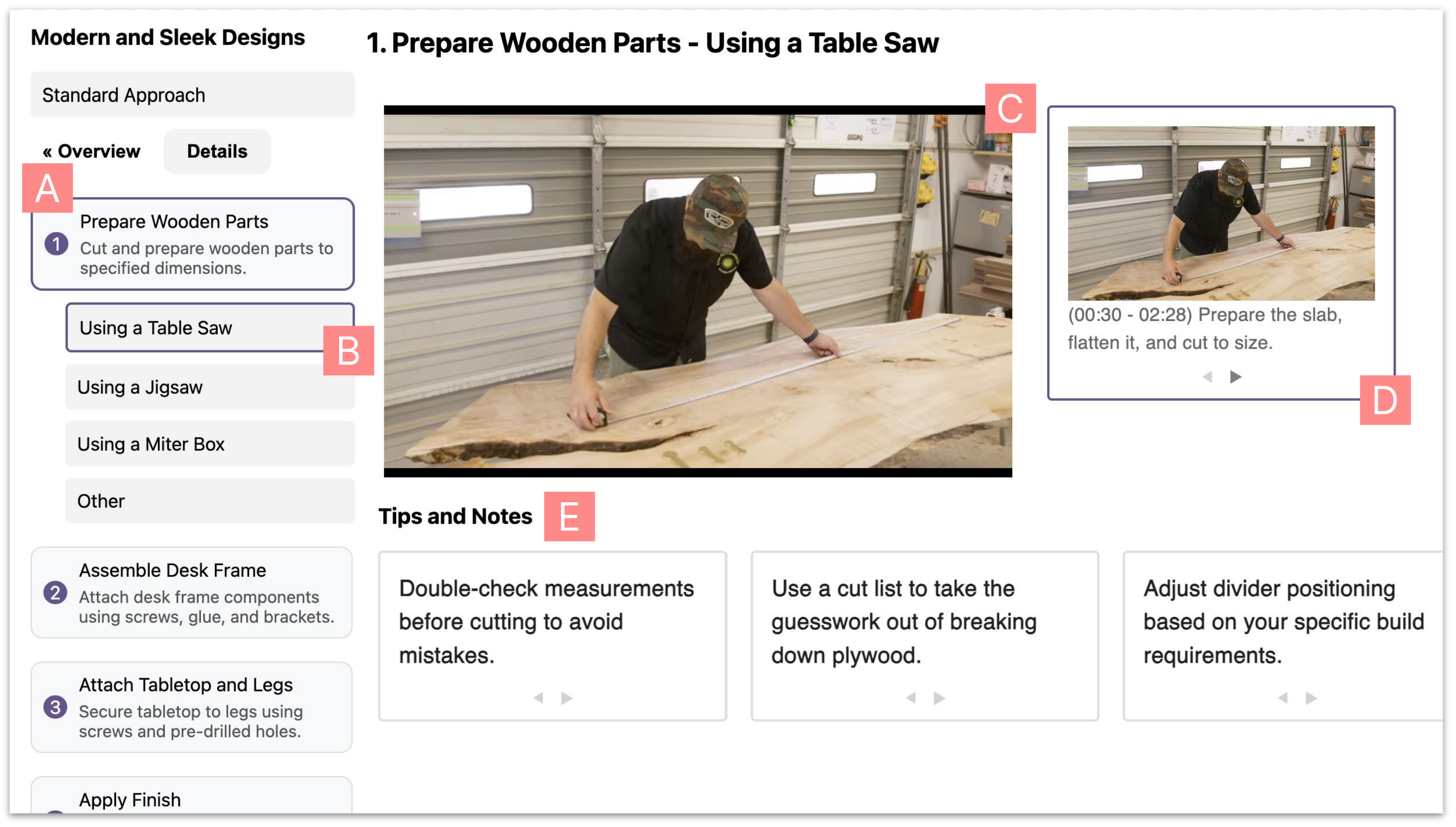}
  \caption{\sysname{} interface on the Details page for the task ``Build a Desk’’. (A) The interface displays the list of steps for the chosen approach. (B) For each step, users can explore different methods, such as tools or techniques, to complete the step. (C) When a method is selected, \sysname{} presents video snippets relevant to that method. (D) Users can easily switch between different videos for the selected method, with the corresponding time frame playing automatically. (E) Additionally, users can view tips and notes extracted from the videos.\protect\footnotemark}
  \label{fig:system-detail}
  \Description{The Details page of VideoMix. There is step information along with methods for each step on the left, and the video player and tips and notes on the right.}
\end{figure*}

In the second phase, we cluster similar outcome descriptions together by outcome types. To extract the outcome type, we first prompt GPT-4o to identify two to four of the most salient themes from the list of video outcome descriptions (Appendix~\ref{sec:prompts_outcome_clustering}). Each theme becomes an outcome type. We then cluster the videos around these outcome types by prompting GPT-4o to assign each video to exactly one outcome type using the video's outcome description (Appendix~\ref{sec:prompts_outcome_assignment}). To provide representative images for each outcome type (Figure ~\ref{fig:system-overview}B), we randomly select two videos assigned to that type. We retrieve the outcome frame segments (identified in phase one) for each video and choose the middle frame of the last segment.

\footnotetext{\rr{Source video: \href{https://youtu.be/fv5bqBehcBc}{youtu.be/fv5bqBehcBc} (Creative Commons licensed)}}

\begin{figure*}[t]
  \includegraphics[width=0.9\linewidth]{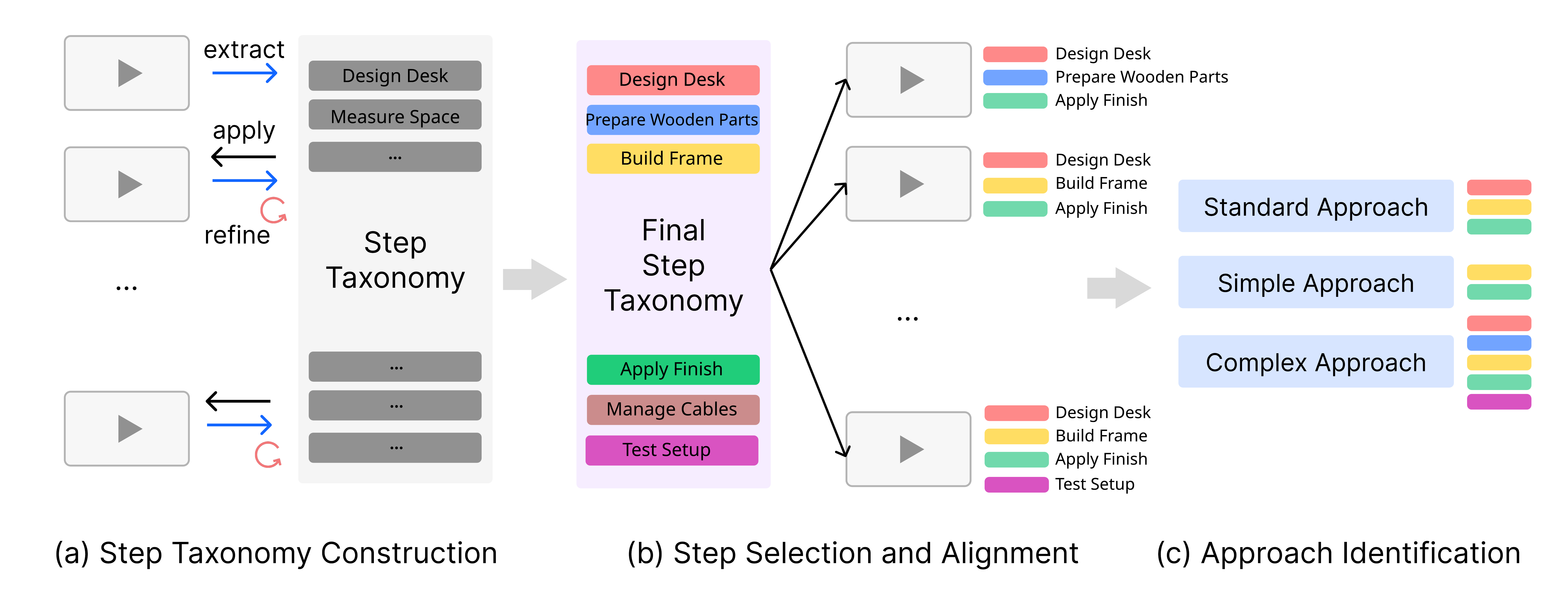}
  \caption{\rr{Illustration of our Dynamic Approach Identification (DAI) module}, which captures a variety of approaches to accomplish a task. (a) The process begins by extracting step information from the first video using GPT-4o. This initial step taxonomy is then applied to the next video, where additional steps are identified, refining the taxonomy. This iterative process continues for all videos, progressively refining the step taxonomy with each comparison. (b) Once the final step taxonomy is established, it is reapplied to each video to detect relevant steps and align segments accordingly. Note that not all steps may be present in each video. (c) After extracting step information from each video using the common taxonomy, the system identifies standard, simple, and complex approaches based on the number of videos that follow each approach and the number of steps within each approach.}
  \label{fig:dai-module}
  \Description{Illustration of the DAI module working in three phases: (a) Step Taxonomy Construction, (b) Step Selection and Alignment, and (c) Approach Identification.}
\end{figure*}

\subsubsection{Steps and Approaches}
To aggregate information from multiple videos sharing the same outcome, it is essential to understand possible sequences of steps that may vary across different videos~\cite{kim23surch}. 
We introduce a \textbf{Dynamic Approach Identification (DAI)} module, which iteratively identifies key steps across a set of videos, accounting for variations in the procedure. Instead of relying on a fixed taxonomy of steps for a task, our module adapts to a specific video pool (in our case, based on the outcome types of the task), and captures procedural differences within the set, ensuring comprehensive coverage of the task.

The DAI module, which is illustrated in Figure~\ref{fig:dai-module}, 
\rr{begins by extracting steps directly from a video transcript and grounding each step in the corresponding transcript sentence indices using GPT-4o (Appendix~\ref{sec:prompts_step_identification}).
Note that prior work~\cite{chen2024tutoai} has demonstrated the feasibility and accuracy of using LLMs for step extraction with timestamps.} 
The extracted step information is then applied to the next video to identify any previously unrecognized steps, adding those new steps to the set. This process is repeated iteratively, refining the step set until the entire video collection is covered. Once the final set of steps (i.e., the final step taxonomy) is derived, the system applies it to each video, selecting the steps present in the video with timestamp information for when each step occurs (Appendix~\ref{sec:prompts_step_assignment}). This method allows us to capture each video’s unique sequence of steps, which may or may not overlap with others.

Once the step information for each video is identified, our pipeline uses the information to determine three approaches: \rr{\textbf{Standard, Simple, and Complex}. The Standard approach refers to the typical sequence of steps most commonly followed across videos. The Simple approach refers to the sequence that involves the fewest steps, while the Complex approach consists of the largest number of steps.}
While there could be other ways to measure the complexity of an approach, we followed Merrill's suggestion~\cite{jobtask} and used the number of steps as a measure, since it provides a quantifiable way to assess the effort required to complete the task.
We execute the process of identifying steps and approaches for each outcome cluster, and the requirements are extracted per each approach. The standard approach is always captured, while the simplest and most complex approaches may not be, particularly if they overlap with the standard approach or if the number of videos following the simplest or most complex approaches is too low. In Section~\ref{sec:tech_eval}, we demonstrate how the DAI module effectively captures diverse and accurate approaches compared to existing baselines.

\subsubsection{Object Requirements}
Our pipeline also extracts the required objects (i.e., the materials, ingredients, and tools used) across all the videos belonging to the same approach. TutoAI~\cite{chen2024tutoai} demonstrated that using LLMs to extract objects from transcripts is the most effective method for identifying items used in tutorial videos. To create a comprehensive list, we also capture \rr{visual frames at 5-second intervals from the entire video}, and together with the entire transcript, prompt GPT-4o to extract the materials, ingredients, and tools used (Appendix~\ref{sec:prompts_requirements}). After gathering this information for each video, we aggregate the results and calculate the frequency of each item across all the videos. To streamline the merging process, we instruct GPT-4o to exclude specific quantities or descriptors (e.g., stripping ``pinch of salt'' to be just ``salt''). 

\subsubsection{Methods and Details}
Finally, our pipeline detects variations in the methods used for each step of an approach. For each step, we get the corresponding transcript segments from all the videos containing that step. We then prompt GPT-4o to identify the different variations in the methods described by transcript segments (Appendix~\ref{sec:prompts_method_clustering}). To identify which of these method variations a video uses, we prompt GPT-4o with the video's step transcript and the method variations and ask it to pick which variation the step transcript describes (Appendix~\ref{sec:prompts_method_assignment}). Finally, for each method, we prompt GPT-4o to extract useful tips or key information by providing a collection of transcript sentences specific to that method (Appendix~\ref{sec:prompts_tips}).

\subsection{Implementation}
The interface for \sysname{} was developed using TypeScript, ReactJS, and CSS. The backend was implemented with Python scripts for video preprocessing. OpenAI’s API was used for VLM components, specifically the GPT-4o-2024-05-13 model~\cite{gpt4o0513} with a temperature setting of 0 for all components. To generate structured outputs, we employed Function Calling~\cite{function_calling} in OpenAI's API. The GPT-4o prompts and their corresponding function calling structure are provided in the Appendix.
\rr{Note that we used GPT-4o to process video frames and transcripts for a robust and scalable solution for handling long-form videos. We did not use video foundation models due to their limited context windows, which make processing lengthy videos challenging without losing details. Future improvements in video foundation models, such as larger context windows and lower costs, could make long-form video processing more efficient and practical for our pipeline.
}

\section{Technical Evaluation}\label{sec:tech_eval}

\begin{table*}[t]
\begin{center}
\renewcommand{\arraystretch}{1.2}
\begin{tabular}{ l c c c c }
 \toprule
 & \multicolumn{3}{c}{\textbf{Accuracy (1-7)}} & \multirow{2}{*}{\textbf{Coverage (0-10)}} \\ 
 & Relevancy & Logical Flow & Completeness &  \\ 
\hline
 \multirow{2}{*}{\textbf{Baseline}} & \multirow{2}{*}{5.88 $\pm$ 1.19} & \multirow{2}{*}{5.58 $\pm$ 1.56} & \multirow{2}{*}{4.50 $\pm$ 1.67} & \multirow{2}{*}{5.80 $\pm$ 2.24}  \\
 & & & & \\
 \hline
 \multirow{2}{*}{\textbf{\sysname{}}} & \multirow{2}{*}{5.42 $\pm$ 1.36} & \multirow{2}{*}{5.52 $\pm$ 1.38} & \multirow{2}{*}{4.42 $\pm$ 1.53} & (1) 7.05 $\pm$ 2.13 (*) \\
 & & & & (2) 7.96 $\pm$ 1.83 (**)\\
 \bottomrule
\end{tabular}
\end{center}
\caption{Results of the technical evaluation of our DAI module. Our pipeline maintained step accuracy across Relevancy, Logical Flow, and Completeness (with no statistically significant differences), while capturing a significantly more diverse range of possible approaches, both (1) when considering only the approaches and (2) across all outcome types (*: p<0.05, **: p<0.01).}
\label{tab:test_table}
\end{table*}

We evaluated the Dynamic Approach Identification (DAI) module primarily, as it is the core component of our pipeline for identifying diverse approaches and methods across multiple videos. We aimed to test two hypotheses: 1) Our pipeline-generated step taxonomy will provide as accurate step information as predefined taxonomies; 2) Our pipeline-generated step taxonomy will better capture the diversity and variation within a task compared to predefined taxonomies.

\subsection{Task Selection}
To evaluate our hypotheses, we selected six tasks from the HowTo100M dataset, with two from the `Hobbies and Crafts' category and four from the `Food and Entertaining' category. The chosen tasks are:
\textit{Build a Desk} (95 videos), \textit{Build a Bookshelf} (58 videos), \textit{Make Chicken Cacciatore} (92 videos), \textit{Make Jambalaya} (66 videos), \textit{Make Shrimp Cocktail} (86 videos), and \textit{Make Bannock} (90 videos). 
The tasks were selected based on the following criteria: 
1) We focused on physical tasks with tangible outcomes, rather than fixing or using products~\cite{yang23howtotaxonomy}. This was to ensure diversity in information, such as outcome types and requirements.
2) The task must have a predefined step taxonomy available in existing datasets (e.g., HT-Step~\cite{Afouras_2023_htstep}, CrossTask~\cite{Zhukov2019crosstask}) to allow for comparison.
3) The task must include at least 50 videos to ensure diversity.
For comparison, we used HT-Step and CrossTask as baseline datasets, since both are also based on HowTo100M. The step taxonomies in these datasets are human-annotated, grounded in WikiHow~\cite{wikihow}, a popular website for how-to instructional articles.

\subsection{Method}
We recruited external evaluators through Prolific~\cite{prolific}, who are familiar with the selected tasks. 
\rr{In total, 24 evaluators were recruited, with 4 evaluators assigned to evaluate each of the 6 tasks.}
To ensure expertise, we required evaluators to self-report having performed the task at least once and to know at least two approaches to completing it. Evaluators were asked to rate the step information derived from both the baseline predefined step taxonomies and our pipeline-generated steps for the same video tasks, where the order of the condition was counterbalanced. 

The evaluation focused on two main criteria following our hypotheses: accuracy and coverage.
For accuracy, evaluators rated the step information based on the following criteria using a 7-point Likert scale:
\begin{itemize}
    \item Relevancy: How relevant is each step to achieving the overall task goal?
    \item Logical Flow: How logical and coherent is the progression of steps in the sequence?
    \item Completeness: How complete is the sequence in covering all necessary steps to achieve the task?
\end{itemize}
For the baseline, evaluators were presented with the predefined step taxonomies, but we summarized each step into a concise step name to ensure consistency with the format of our pipeline-generated steps. For our pipeline-generated taxonomies, evaluators were provided with the `standard' approach for each outcome type.
For coverage,  evaluators answered the following question on a scale of 0 to 10 where 0 indicates no coverage and 10 means a full, 100\% coverage:
\begin{itemize}
    \item  To what extent does this sequence represent or cover all the possible ways to achieve the task? 
\end{itemize}
In this case, evaluators were provided not only with the standard approach but also with simple and complex approaches for each cluster, if available, in the pipeline-generated taxonomies. Evaluators were compensated \$5 USD for each task they evaluated, which took approximately 15 minutes.

\subsection{Results}
Overall, our pipeline maintained step accuracy while capturing a more diverse range (80\%) of possible approaches compared to the baseline (58\%).
For accuracy, when evaluated on three key aspects---Relevancy, Logical Flow, and Completeness---using a 7-point Likert scale, there were no statistically significant differences between the steps generated by our pipeline and those annotated by humans.
(Table~\ref{tab:test_table}, Relevancy: $\mu$=5.88, $\sigma$=1.19 vs. $\mu$=5.42, $\sigma$=1.36; $Z$=1.5, $p$=0.13,
Logical Flow: $\mu$=5.58, $\sigma$=1.56 vs. $\mu$=5.52, $\sigma$=1.38; $Z$=0.37, $p$=0.71,
Completeness: $\mu$=4.5, $\sigma$=1.67 vs. $\mu$=4.42, $\sigma$=1.53; $Z$=0.47, $p$=0.64).
Note that each condition was evaluated according to its intended outcome. The baseline involved the general task (e.g., building a desk), while our pipeline was tested on specific outcomes (e.g., building a standing adjustable desk). These results indicate that our pipeline can generate steps with a level of quality comparable to human-annotated steps, even when addressing more specific tasks.

In terms of coverage, the steps generated by our pipeline captured a significantly greater range of possible approaches to completing the task, as rated on a scale from 0 to 10, (0 being 0\% and 10 being 100\%). Compared to the baseline steps, our pipeline captured a more diverse range of approaches, even when considering only the approaches (i.e., Standard, Simple, and Complex) for each intended outcome type. 
(Table~\ref{tab:test_table}, $\mu$=5.8, $\sigma$=2.24 vs. $\mu$=7.05, $\sigma$=2.13; $Z$=-2.16, $p$<0.05).
When aggregating these approaches across all outcome types, the coverage increased significantly from 58\% to 80\%, with an average of 3.5 outcome types per task 
(Table~\ref{tab:test_table}, $\mu$=5.8, $\sigma$=2.24 vs. $\mu$=7.96, $\sigma$=1.83; $Z$=-3.37, $p$<0.01). 
These results demonstrate that our pipeline, which detects step information across various outcome types and approaches, captures significantly more diverse ways to achieve a task. All statistical significance was measured using the Wilcoxon Rank-Sum Test.

\section{User Study}

We conducted a within-subjects user study to evaluate \sysname{} against a baseline YouTube-like system, a platform most users are familiar with for watching how-to videos. The primary goal of the study was to assess the effectiveness of \sysname{} in enhancing users’ overall understanding of tasks, and to explore how users would use \sysname{} and how it impacts their learning experience.

\subsection{Participants and Apparatus}
We recruited 12 participants (4 male, 8 female, \rr{mean age=25.3, median=25.5}) through an online
community at our academic institution, those who regularly watch how-to videos and often watch multiple videos to learn a specific task. 
For the study, we selected 4 tasks from those used in our pipeline evaluation: two from the `Hobbies and Crafts' category (\textit{Build a Desk, Build a Bookshelf}), and two from the `Food and Entertaining' category (\textit{Make Chicken Cacciatore, Make Jambalaya}). We randomly selected two tasks for each participant, one for \sysname{} and the other for a baseline system.
Since our study involved learning tasks, we ensured that none of the participants had prior experience with the tasks they would be learning during the session.

For a fair comparison between \sysname{} and baseline, we built a baseline system similar to YouTube, but with a limited set of videos available in \sysname{}. Participants were provided with a list of videos in the main feed, where they could click to watch each video along with its title and description sourced from the original YouTube video.

\subsection{Study Procedure}
The study was conducted online through a Zoom meeting. Participants were first given an overview of the study, including the two tasks they would be learning during the study. They were then instructed to use either \sysname{} or the baseline system to learn about an assigned task. Participants were asked to imagine they would later be performing the task on their own, and that their current goal was to study the task, gather as much information as possible to prepare for it.

We provided a brief tutorial on how the assigned system worked, and participants were given 15-20 minutes to explore and learn about the task using the system. 
They were encouraged to think aloud, sharing their thoughts and decision-making process as they use the systems. After completing one session, participants switched to the other system, and the same process was repeated. The order of tasks and systems used were counterbalanced across participants. 
Following each session, we conducted a questionnaire to assess participants' perceived understanding of the task, perceived usefulness of each feature (in the \sysname{} condition only), and cognitive load using measures from NASA-TLX (\textit{Mental Demand, Frustration, Effort, Performance})~\cite{hart1988nasatlx}. 
All responses were on a 7-point Likert scale. Finally, we conducted semi-structured interviews to understand their strategies used in each system and gather qualitative feedback on \sysname{}.
The study lasted 1 hour, and participants were compensated with a \$30 USD Amazon gift card.

\subsection{Results}

\begin{figure}[t]
  \includegraphics[width=\linewidth]{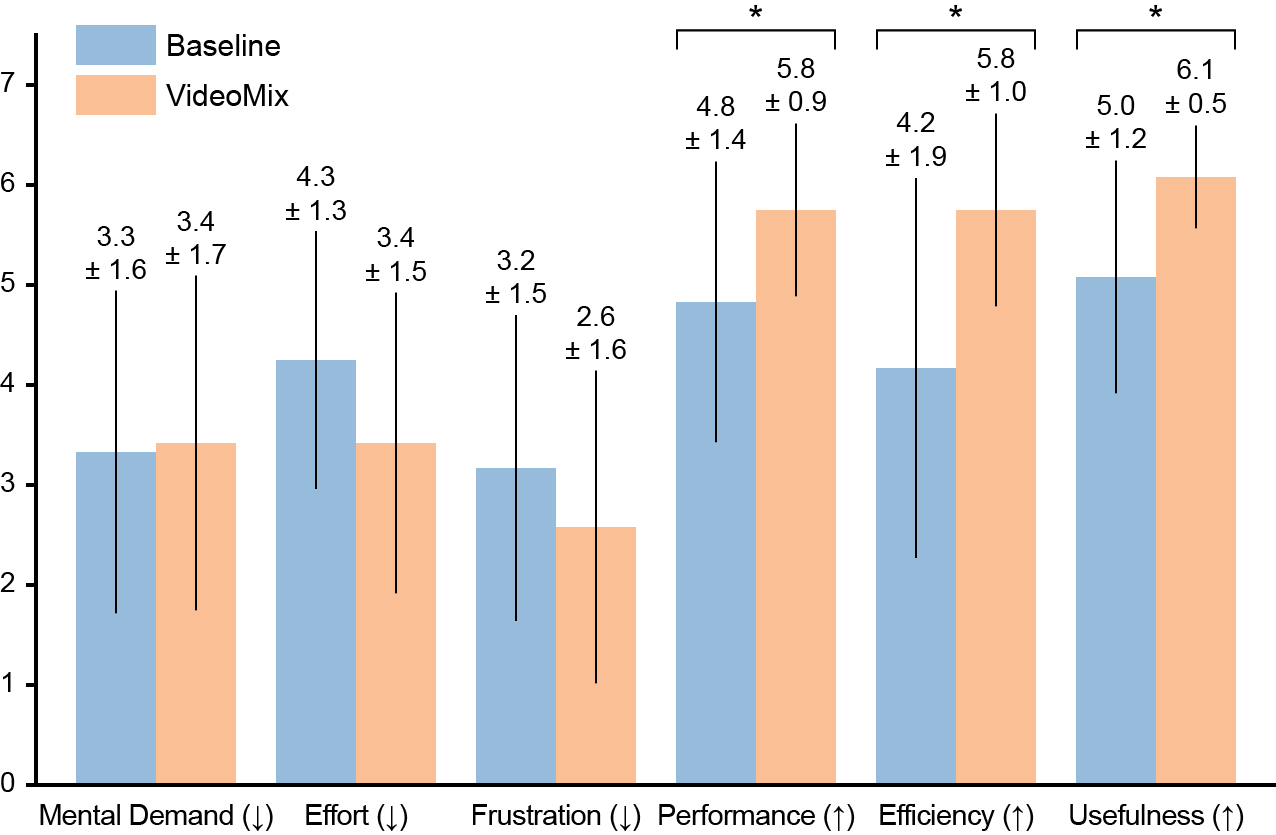}
  \caption{Participants felt they were more successful and efficient with \sysname{}, and found \sysname{} to be more useful when learning about the task compared to the baseline. There were no statistically significant differences in mental demand, effort, and frustration (*: p<0.05).}
  \label{fig:user-study-comparison}
  \Description{Bar chart comparing user experiences in Baseline and VideoMix}
\end{figure}

Overall, participants found \sysname{} to be more helpful in understanding the task compared to the baseline. Below, we provide a detailed report of the study’s findings. For all measures, we first conducted a Shapiro-Wilk test to determine data normality, and then used a paired t-test (if parametric) and a Wilcoxon signed-rank test (if non-parametric).

\begin{figure*}[t]
  \includegraphics[width=0.9\linewidth]{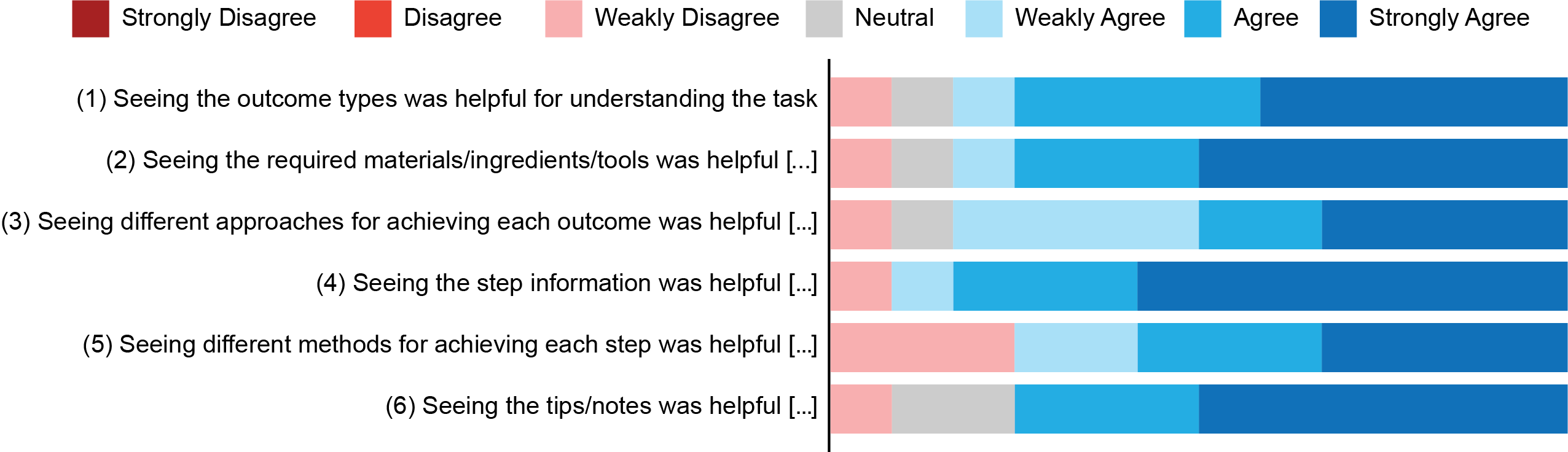}
  \caption{Participants' ratings on the usefulness of each information piece in understanding the task. Overall, they found the information provided by \sysname{}—including outcome types, requirements, different approaches, step details, methods, and tips and notes—to be helpful in gaining a better understanding of the task.}
  \label{fig:system}
  \Description{Bar chart showing survey responses on VideoMix}
\end{figure*}

\subsubsection{Enhanced Overall Understanding}
Participants reported a significantly better understanding of the tasks when using \sysname{} compared to the baseline (Figure~\ref{fig:user-study-comparison}). They felt more successful in learning about the task ($\mu$=4.83, $\sigma$=1.4 vs. $\mu$=5.75, $\sigma$=0.83; $t$=-2.42, $p$<0.05) and more efficient in the learning process ($\mu$=4.17, $\sigma$=1.9 vs. $\mu$=5.75, $\sigma$=0.92; $W$=7.0, $p$<0.05). 
Participants appreciated how \sysname{} provided a comprehensive overview of the task, allowing them to grasp the entire scope at a glance. For instance, P2 noted, \textit{``With \sysname{}, I could see the overall process involved in the task and get a general understanding immediately. I could figure out possible outcomes, required materials, and overall process, which would have taken a long time to find on YouTube, where videos are scattered.''}

\sysname{} significantly streamlined the process of acquiring task-related information compared to the baseline. With the baseline, participants typically relied on thumbnails to identify the outcome or titles to see the approach they wanted (e.g., `simple recipe').
After selecting a video, they would check the description box in hopes of finding a list of ingredients or basic step-by-step instructions, but this information was not always available. In contrast, \sysname{} offered organized information upfront, saving participants considerable time.  For example, P6 selected the standard approach in \sysname{} because he wanted to learn something basic, whereas on the baseline, he had to watch multiple videos and compare processes to identify the original standard recipe. He also mentioned, \textit{``It's nice because the ingredients are written out, so you can just look at them and prepare everything right away.''} While \sysname{} presented information from an average of 77.8 videos per task, participants watched only 2.6 videos on average using the baseline system within the given study time.

Overall, participants rated \sysname{} to be significantly more useful for gaining an overall understanding of the task compared to the baseline system ($\mu$=5.08, $\sigma$=1.16 vs. $\mu$=6.08, $\sigma$=0.49; $t$=-2.87, $p$<0.05 ). 
10 out of 12 participants mentioned they would prefer \sysname{} to baseline when understanding a task. 
However, there were no statistically significant differences in mental load, frustration, or effort during the learning process.

\subsubsection{Tailored Learning Experience}
VideoMix organizes instructional content from multiple videos into a hierarchical structure based on outcome, approach, and method employed. This allowed participants to efficiently focus on instructions that best suited their specific needs and context.

First, the outcome types helped participants narrow their focus to what they were most interested in learning. After exploring various outcome choices,  participants developed a clear preference based on either personal tastes (e.g., \textit{Jambalaya with Chicken and Sausage} vs. \textit{Vegan or Low-Carb Jambalaya}) or estimated proficiency level (e.g., \textit{Modern and Sleek Design Desk} vs. \textit{Functional and Multi-purpose Design Desk}).

Next, the different approaches enabled participants to choose learning pathways that matched their experience level. Most participants, being new to the task, looked for simple or standard methods. P8 remarked, \textit{``It was easy to have a clear criterion, whereas on YouTube, I had to guess content from thumbnails and titles. Even if the first video I watched had a unique approach, I might have assumed it was the original recipe for Jambalaya. I would have spent much more time than I did with VideoMix to find a recipe that fit my situation.''} P8 only realized that one of the three videos she watched on YouTube matched her beginner level after viewing all three.

Finally, the variety of methods allowed participants to focus on instructions that aligned with their available tools and ingredients. For example, P3 said, \textit{``It was helpful to see different methods because I don’t have an oven, so I looked at the \texttt{Using Stove} or \texttt{Using Pot} methods instead of \texttt{Using Oven}.''} In contrast, finding a video that fit their context on YouTube was often more challenging. P7 noted, \textit{``As I watched the video, I was concerned that I didn’t have the right equipment or materials used in the video, and  thought I’d probably need to search for another one.''}
In summary, VideoMix enabled participants to learn more effectively by providing clear, relevant options that could be tailored to their specific preferences and resources.

\subsubsection{Knowledge Acquisition By Multi-Video Comparison}\label{sec:user-study-knowledge}

\sysname{} allows users to easily navigate between videos showcasing the same method within a step (Figure~\ref{fig:system-detail}D). By comparing multiple segments, participants gained a deeper understanding of the methods. For example, P4 said, \textit{``Even though both video segments I watched were all about using wood glue, one video showed how to apply it while the other explained when to use it. This helped me understand the step better.''} Similarly, P1 initially didn’t know what Leger Boards were in Building a Bookshelf when only watched a single video segment, but learned what they are after watching multiple segments using them.

Participants also picked up key information about requirements and techniques. For example, P8 said, \textit{``I saw that celery and garlic were used across all standard approaches of different outcomes, so I realized they are key ingredients.''} P9 highlighted how different methods offered contrasting tips, saying, \textit{``For the Instant Pot, tips suggested adding vegetables first, while for the rice cooker, they recommended adding meat first. The order seems important based on the tool you're using.''} The ability to compare multiple perspectives within the same task participants' understanding, offering a more comprehensive learning experience.

\subsubsection{Further Improvements for \sysname{}}\label{sec:user-study-feedback}
While participants found \sysname{} to be an effective tool for learning new tasks through videos, they noted suggestions on how \sysname{} can be further improved. First, they mentioned the discontinuous nature of the segmented videos throughout the steps could hinder the learning process. 
For example, P1 said, \textit{``When I clicked a next step and a segment from a new video was shown, it took me a while to understand the context of the video.''}
Participants expressed a desire to see a continuous video, while having the information \sysname{} offers. P2 said, \textit{``It would be great if I could select one main video, and see additional details not covered in that video through \sysname{}.''} A potential improvement could be a hybrid format, where users first watch a full video, and what VideoMix currently provides is organized around that primary video.

Participants also suggested ideas on how methods are presented. For example, P10 suggested sorting the methods by commonness, similar to how requirements are organized or how \sysname{} shows the `Standard Approach' (as we do for the approaches). P12 wished to see the outcome of each video segment to better choose which method to follow, similar to how \sysname{} shows different outcome types for the task on the overview page. This feedback suggests that the hierarchical structure \sysname{} uses to organize task-level information could be re-applied at the step level, providing more detailed information.

\section{Discussion}
In this paper, we present \sysname{}, a system that aggregates multiple how-to videos to provide a comprehensive understanding of a task. We discuss how it supports task learning, considerations for designing multi-video systems, the incorporation of the hierarchical nature of tasks, and potential directions for future work.

\subsection{Supporting Task Learning: from Understanding to Following}\label{sec:discussion-task}

\rr{\sysname{} is designed to facilitate task learning by helping users synthesize multiple videos, enabling a better understanding of the task. This aligns with a key search intention in Information Retrieval~\cite{shah22search}, which emphasizes learning domain knowledge. While \sysname{} is primarily intended to assist users in the understanding phase before they move on to task execution, 7 out of 12 participants expressed interest in using it throughout the task-following phase as well.}

Participants highlighted several benefits of \sysname{} in task following: it presents various methods together, allowing users to choose their preferred approach as they follow the task without searching through multiple videos (P2); its mixed-media format with text makes following instructions easier (P6); and the segmented steps enable users to quickly revisit specific parts of the process (P11). 

However, other participants preferred YouTube for task \textit{following}, citing the importance of consistency and flow. As P9 noted, \textit{``Mixing two different recipes is generally not a good idea.''}
While a few participants could identify the same video across different steps by recognizing the background or demonstrator, \rr{it remains important to support the tracking of a cohesive procedure within a single video, especially in the following phase.}

To better support the full learning cycle---from understanding to following---we envision a system that allows users to explore various methods (as \sysname{} currently does), then select specific videos for following, while maintaining easy access to overview information ~\cite{kim23surch}. 
\rr{To better support the following phase, we suggest features like real-time prompting or interactive search to address the users’ more specific needs as they progress through the task. Additionally, while \sysname{} offers some customization by providing a list of tools for each approach and outcome type, or methods specifying tool usage, allowing users to retrieve videos based on selected tools or choose the level of detail they want to explore could further improve the customization experience.}


\subsection{Designing Multi-Video Systems}
\sysname{} organizes information from multiple videos to provide a comprehensive understanding of tasks. Instead of treating videos as the primary object, \sysname{} treats the task itself as the first-class object, with multiple videos structured around it. Thus, the basic unit is a video segment (i.e., part of a video), which is then organized around a task.

Designing a multi-video interface around video segments has both advantages and challenges. On the positive side, splitting content by steps made it easier to digest, and multiple demonstrations for each step enhanced learning (Section~\ref{sec:user-study-knowledge}). However, users could feel a sense of discontinuity between segments and sometimes lack the broader context of the full video (Section~\ref{sec:user-study-feedback}). To address this, a multi-video interface should ensure that enough context is provided and consider strategies to maintain continuity, such as using a common voice-over, visual connectors, or a consistent theme across videos.

\rr{Another challenge is managing the extensive amount of information drawn from multiple videos, which may feel overwhelming to some users. Two out of 12 participants who preferred YouTube over \sysname{} appreciated its ability to present diverse methods at a glance but found the overall information density to be excessive. While \sysname{} aims to reduce the time required to learn viable methods through structured presentation—particularly for tasks with high variability—this comes with trade-offs. Curating information may limit certain details as well, and it is essential to balance organization with user agency in the exploratory search process.}

Lastly, it would be interesting to explore how a multi-video interface might reshape user engagement, especially in interactions typically supported by traditional video-centered platforms, such as commenting, liking, or sharing. Investigating how these interactions can be adapted to a multi-video interface, as well as identifying potential new interactions unique to this interface, would be an interesting avenue for future research.

\subsection{Incorporating Hierarchical Nature of Tasks}\label{sec:discussion-hierarchy}

How-to videos naturally contain hierarchical information~\cite{Zala2023HiREST}. Tasks often consist of multiple sub-tasks or steps, each of which could be a task on its own. For example, in the task of making an Eggs Benedict, one of the steps might involve poaching an egg, where there could be videos solely about it.

This hierarchical structure presents an opportunity for \sysname{} to further enhance learning by extending its current task-level organization to a more granular, step-level structure. Just as \sysname{} organizes information by outcome, requirements, and approaches at the task level, the same principle could be applied recursively at the step level (as briefly discussed in  Section~\ref{sec:user-study-feedback}). For instance, the step of poaching an egg could be broken down into sub-steps such as preparing water, cracking the egg, and cooking the egg, where there could be multiple variations within each sub-step. This approach would allow users to delve deeper into specific areas of interest, fostering a more flexible and personalized learning experience. By supporting this recursive exploration, users could not only learn how to complete a task like making an Eggs Benedict but also master individual skills, like poaching eggs, that could be applied in a wide range of other contexts, supporting a flexible and infinite journey of learning.

\subsection{Limitations and Future Work}
\rr{Our pipeline only requires videos to have narration, as it relies on spoken content to extract task steps and details. As long as videos are accessible and can be transcribed using ASR, our approach remains applicable.}
However, a key limitation of \sysname{} is its dependence on the quality and quantity of the source videos. Since the system compiles content from various videos, the clarity of the presenter’s instructions and the logical flow of the content can significantly affect its performance. In particular, \sysname{} relies heavily on transcripts for extracting steps and methods, making clear and well-structured narration essential. If a video lacks coherence or clarity, the system may struggle to extract accurate and meaningful information.

In terms of quantity, our system may not provide as comprehensive an overview when the available videos are limited (e.g., only 10 videos on a given task). In such cases, we could consider expanding the search to include more videos (e.g., similar methods used in different tasks) or incorporating other tutorial resources, such as text-based materials.
Similarly, while we demonstrated \sysname{} based on videos selected from the HowTo100M dataset~\cite{miech19howto100m}, expanding the video pool through additional crawling would allow \sysname{} to offer a richer and more diverse set of instructions. By refining search queries to capture more hierarchical videos (e.g., searching for specific outcome clusters or individual methods), the system could provide a broader range of instructional content. We believe that as \sysname{} processes more videos, its comprehensiveness and ability to support users will improve.

Additionally, \sysname{} has primarily been tested on tasks involving the creation of physical objects, which typically feature well-defined steps and clear visual and verbal cues. However, extending \sysname{} to other types of tasks---such as digital tasks like Photoshop editing or guitar tutorials---may introduce new challenges. For example, tasks like guitar tutorials may require a different structure that emphasizes progressive skill building rather than multiple methods to achieve the same step. They may also rely more heavily on subtle nuances such as hand placement, tone, or timing, which are difficult to capture solely through transcripts.
Beyond how-to tasks, there is potential for \sysname{} to be applied to other domains, such as organizing interview videos by specific questions or themes. By structuring interviews around common topics across multiple videos, the system could provide users with a comprehensive view of diverse perspectives. This approach could also be extended to educational content, where \sysname{} could organize lectures by subtopics, offering a clearer, more structured learning path for users.

\section{Conclusion}
This paper presents \sysname{}, a system that helps users gain a comprehensive understanding of how-to tasks by aggregating information from multiple tutorial videos. We demonstrated that \sysname{} enables users to explore different methods, materials, and outcomes more easily, leading to a better understanding of a task.
Our work highlights the potential of a task-oriented, multi-video approach to support users in task learning. As online tutorials and video content continue to grow, our system provides an important step forward in improving how people learn from them.

\begin{acks}
\rr{We thank Kim Pimmel and Mira Dontcheva for their feedback on our early research  prototype design.
The videos and screenshots used in our paper are licensed under CC BY 3.0, including \href{https://youtu.be/CbJtZFXwxKY}{\textit{W101\_Corner desk for drawing}} by \textit{ViTTEN}, \href{https://youtu.be/Fnl1OwAAvEo}{\textit{Build an easy desk Under \$50}} by \textit{Abdullah's informative channel}, \href{https://youtu.be/Z7x_Rvb_yjc}{\textit{Cheap and Simple DIY Farmhouse Desk Build}} by \textit{Kim and Garrett Make It!}, \href{https://youtu.be/_v0fXgwcrpY}{\textit{TILE TOP COMPUTER TABLE with PLANS}} by \textit{D.A Santos}, and \href{https://youtu.be/fv5bqBehcBc}{\textit{A Woodworkers Dream Desk}} by \textit{Coffey Custom Builds}.
This work was supported by the Institute of Information \& Communications Technology Planning \& Evaluation (IITP) grant funded by the Korean government (MSIT) (No.2021-0-01347, Video Interaction Technologies Using Object-Oriented Video Modeling).}
\end{acks}

\bibliographystyle{ACM-Reference-Format}
\bibliography{main}

\appendix

\section{GPT Prompts - (1) Outcomes}\label{sec:prompts_outcome}
\subsection{Transcript About Outcomes}\label{sec:prompts_outcome_transcript}

\begin{lstlisting}[escapechar=\%, breaklines=true, breakatwhitespace=true]
You are given the transcript of the tutorial video related to {task_name}.
Identify the part of the transcript that describes the fnal results of the procedure, such as "Look
how beautiful our cake turned out."
The transcript is as follows: {transcript_data}
Return the sentence indices that describe the outcome.
"""

\end{lstlisting}

\begin{lstlisting}[escapechar=\%]
"parameters": {
    "type": "object",
    "properties": {
        "index": {
            "type": "array",
            "items": {
                "type": "integer"
            }
        }
    },
    "required": ["index"]
}
            
\end{lstlisting}

\subsection{Outcome Descriptions}\label{sec:prompts_outcome_description}
\begin{lstlisting}[escapechar=\%, breaklines=true, breakatwhitespace=true]
{visual_frames}
The transcript of this tutorial video related to {task_name} is as follows: {transcript_data}.
Provide a one-sentence description of the final outcome of the tutorial video related to {task_name}. 
Focus solely on its appearance without introductory phrases, subjective language, or references to the methods used.
"""

\end{lstlisting}

\subsection{Outcome Clustering}\label{sec:prompts_outcome_clustering}
\begin{lstlisting}[escapechar=\%, breaklines=true, breakatwhitespace=true]
You are provided with descriptions of outcomes from tutorial videos related to {task_name}.
Your task is to group similar outcomes into clusters based on common themes. 
Create between 2 and 4 clusters, and assign a descriptive name to each cluster that reflects the shared theme of the outcomes within it. Return only the names of the clusters.

Below are the outcome descriptions from each video:
{outcome_descriptions}
"""

\end{lstlisting}

\begin{lstlisting}[escapechar=\%, breaklines=true, breakatwhitespace=true]
"parameters": {
    "type": "object",
    "properties": {
        "clusters": {
            "type": "array",
            "items": {
                "type": "string"
            },
        }
    },
    "required": ["clusters"]
}
            
\end{lstlisting}

\subsection{Outcome Assignment}\label{sec:prompts_outcome_assignment}
\begin{lstlisting}[escapechar=\%, breaklines=true, breakatwhitespace=true]

Assign the video outcome to one of the outcome types.
video outcome: {outcome_description}
outcome types: {outcome_types}
"""

\end{lstlisting}

\begin{lstlisting}[escapechar=\%, breaklines=true, breakatwhitespace=true]
"parameters": {
    "type": "object",
    "properties": {
        "outcome": {"type": "string", "enum": outcome_type_list},
    },
    "required": ["outcome"]
}
            
\end{lstlisting}

\section{GPT Prompts - (2) Requirements}\label{sec:prompts_requirements}

\begin{lstlisting}[escapechar=\%, breaklines=true, breakatwhitespace=true]
{visual_frames}
The transcript of this tutorial video related to {task_name} is as follows: {transcript_data}.

Identify the ingredients, tools and equipment used in this tutorial video.\n
Extract and list the ingredients, tools, and equipment without specifying quantities or any descriptors.
"""

\end{lstlisting}

\begin{lstlisting}[escapechar=\%, breaklines=true, breakatwhitespace=true]
"parameters": {
    "type": "object",
    "properties": {
        "ingredients": {
            "type": "array",
            "items": {
                "type": "string"
            }
        },
        "tools and equipment": {
            "type": "array",
            "items": {
                "type": "string"
            }
        }
    },
    "required": ["ingredients", "tools and equipment"]
}
            
\end{lstlisting}

\section{GPT Prompts - (3) Steps}\label{sec:prompts_steps}

\subsection{Step Identification}\label{sec:prompts_step_identification}
\begin{lstlisting}[escapechar=\%, breaklines=true, breakatwhitespace=true]
Given the transcript of a tutorial video related to {task_name}, extract the key high-level steps involved in the task.
Follow these guidelines when extracting steps:
1. Steps should be high-level and concise.
2. Base each step on an intermediate outcome with tangible results (e.g., "Make Dough", "Grill Steak"), instead of individual actions (e.g., "Add Flour", "Turn on Grill").
3. Avoid using specific ingredients in the step name (e.g., "Add Tomato Paste"). Instead, focus on the purpose of the step (e.g., "Make Sauce" instead of "Add Tomato Paste").
4. Group together related low-level actions into a single, high-level step. (e.g., combine "Add Salt" and "Add Lime" into "Make Sauce").
5. A step must span multiple transcript sentences, not just a single sentence. It should be high-level enough.
6. Use a concise "verb + object" format to describe each step, containing only one verb (e.g., "Boil Potatoes").
7. Exclude any steps unrelated to the core task, such as introductions, conclusions, or general commentary.

First, review the existing list of steps to identify if any of them are mentioned in the transcript. 
Use the same step names to ensure consistency whenever possible.
If you identify new steps that are not in the existing list, add them appropriately.

Here is the existing list of steps:
{original_step}

Here is the transcript of the videos:
{transcript_data}

Return a series of concise, high-level steps as a list.
"""

\end{lstlisting}

\begin{lstlisting}[escapechar=\%, breaklines=true, breakatwhitespace=true]
"parameters": {
    "type": "object",
    "properties": {
        "steps": {
            "type": "array",
            "items": {
                "type": "string"
            },
        }
    },
    "required": ["steps"]
}
\end{lstlisting}

\subsection{Step Assignment}\label{sec:prompts_step_assignment}
\begin{lstlisting}[escapechar=\%, breaklines=true, breakatwhitespace=true]
You are provided with a transcript of a tutorial video about {task_name}, along with a list of possible steps for the task.

Your task is to read through the transcript sequentially and assign the appropriate steps from the provided list to the corresponding sections of the transcript. 
The steps may not be in order in the list, and some steps may not be used at all. Only assign a step when the content clearly matches the step.
For each step you assign, specify the corresponding section of the transcript by identifying the start and end indices. 

Here is the list of steps:
{whole_step}

Here is the transcript data:
{transcript_data}

Return the assigned steps in the order they occur in the transcript.
"""

\end{lstlisting}
\begin{lstlisting}[escapechar=\%, breaklines=true, breakatwhitespace=true]
 "parameters": {
    "type": "object",
    "properties": {
        "steps": {
            "type": "array",
            "items": {
                "type": "object",
                "properties": {
                    "step_name": {"type": "string"},
                    "sentence_start": {"type": "integer"},
                    "sentence_end": {"type": "integer"}
                },
                "required": ["step_name", "sentence_start", "sentence_end"]
            }
        }
    },
    "required": ["steps"]
}
\end{lstlisting}

\section{GPT Prompts - (4) Methods and Tips}\label{sec:prompts_methods}

\subsection{Method Clustering}\label{sec:prompts_method_clustering}
\begin{lstlisting}[escapechar=\%, breaklines=true, breakatwhitespace=true]
You are given video transcripts from multiple videos about {task_name}, all demonstrating the same step, "{step_name}". 
Your task is to cluster the different methods or approaches used in these videos.
When clustering, focus on the type of tools, equipment, or techniques used.

Examples:
- Step: Boil Potatoes
-- Variations: Boiling using stove, Boiling using oven, Boiling using microwave.

- Step: Mix Ingredients
-- Variations: Mixing with spoon, Mixing with whisk, Mixing with blender.

Remember to ground variations on the provided video transcripts.
To ensure clustering based on the same criteria, each cluster name should start with the same action word (e.g., "Using [Tool Name]", "Applying [Technique]").
Create up to 3 clusters based on these variations.

Here are the transcripts of the step: 
{transcripts}
"""

\end{lstlisting}

\begin{lstlisting}[escapechar=\%, breaklines=true, breakatwhitespace=true]
"parameters": {
    "type": "object",
    "properties": {
        "clusters": {
            "type": "array",
            "items": {
                "type": "string"
            },
        }
    },
    "required": ["clusters"]
}
\end{lstlisting}

\subsection{Method Assignment}\label{sec:prompts_method_assignment}
\begin{lstlisting}[escapechar=\%, breaklines=true, breakatwhitespace=true]
You are given a video transcript demonstrating the step "{step_name}". 
Your task is to assign the method described in the transcript to one of the existing method types.

Method types: {variation}
Video transcript: {transcript}

Identify which method type best matches the approach described in the transcript.
"""

\end{lstlisting}
\begin{lstlisting}[escapechar=\%, breaklines=true, breakatwhitespace=true]
 "parameters": {
    "type": "object",
    "properties": {
        "method": {"type": "string", "enum": method_type_list},
    },
    "required": ["method"]
}
\end{lstlisting}

\subsection{Tip Extraction}\label{sec:prompts_tips}
\begin{lstlisting}[escapechar=\%, breaklines=true, breakatwhitespace=true]
 You are provided with video transcripts about {task_name}, focusing on the part related to {step_name}. 
Your task is to extract useful tips from the transcripts.
Tips can include:

- Suggestions to improve efficiency or results
- Common mistakes to avoid
- Best practices to follow
- Warnings or important reminders

Extract the top 3 most common tips, advice, or recommendations from the transcripts. 
You should ground each tip on the sentence indices in the transcript where they were found.
You can include more than one video for each tip if the tip is mentioned in multiple videos.

Transcripts: {transcripts}
"""

\end{lstlisting}
\begin{lstlisting}[escapechar=\%, breaklines=true, breakatwhitespace=true]
"parameters": {
    "type": "object",
    "properties": {
        "tips": {
            "type": "array",
            "items": {
                "type": "object",
                "properties": {
                    "tip": {
                        "type": "string"
                    },
                    "videos": {
                        "type": "array",
                        "items": {
                            "type": "object",
                            "properties": {
                                "video_id": {
                                    "type": "string"
                                },
                                "start_index": {
                                    "type": "number"
                                },
                                "end_index": {
                                    "type": "number"
                                },
                            },
                            "required": ["video_id", "start_index", "end_index"]
                        }
                    }
                },
                "required": ["tip", "videos"]
            },
        }
    },
    "required": ["tips"]
}
\end{lstlisting}

\end{document}